\documentclass[11pt,conference]{IEEEtran}
\usepackage{cite}
\usepackage{amsmath,amssymb,amsfonts}
\usepackage{algorithmic}
\usepackage{graphicx}
\usepackage{textcomp}
\usepackage{xcolor}

\usepackage{dblfloatfix}
\usepackage{tikz}
\usetikzlibrary{arrows.meta, positioning, shapes.geometric, fit, backgrounds, calc}
\usepackage{algorithm}
\usepackage{algorithmic}
\usepackage{subcaption}
\usepackage{cite}
\def\BibTeX{{\rm B\kern-.05em{\sc i\kern-.025em b}\kern-.08em
    T\kern-.1667em\lower.7ex\hbox{E}\kern-.125emX}}
\begin{document}

\title{A Structured State Space Sequence Model for Multi-Class Classification of Malware}

\author{\IEEEauthorblockN{Emmanuela Andam\textsuperscript{1,2}, Rana Shaaban\textsuperscript{2}, Emanuel Grant\textsuperscript{2}, and Naima Kaabouch\textsuperscript{1,2}}
\IEEEauthorblockA{\textsuperscript{1}Artificial Intelligence Research (AIR) Center\\
\textsuperscript{2}School of Electrical Engineering and Computer Science, College of Engineering \& Mines\\
University of North Dakota, Grand Forks, ND 58201, USA}}

\maketitle

\begin{center}
\footnotesize
\textcopyright~2026 IEEE. Personal use of this material is permitted. Permission from IEEE must be obtained for all other uses, in any current or future media, including reprinting/republishing this material for advertising or promotional purposes, creating new collective works, for resale or redistribution to servers or lists, or reuse of any copyrighted component of this work in other works.
\end{center}

\begin{abstract}
By 2030, Internet of Things (IoT) devices are projected to reach 40 billion, with fast-paced technological advancements in fields such as industry, healthcare, agriculture, automobiles, and building/home automation systems. This expansion has created a large attack surface for cybercrime, as the majority of these devices open the door for cybercriminals to exploit vulnerabilities, as they lack adequate built-in security. Cybercriminals launch malware attacks to compromise systems or steal sensitive data, and once a system is compromised, a ransom is typically demanded for its release. Current cybersecurity measures in place are being outpaced by the rapid growth of the IoT, which is accompanied by a subsequent growth in malware variants being created per day. Recognizing this pitfall, this research examines and proposes a novel approach to malware detection and classification to safeguard devices from further attacks and make IoT systems more robust and secure. The framework proposed utilizes a Structured State Space Sequence (S4) model, which discretizes sequences of malware samples in a sequence and captures long-range dependencies, essentially identifying the “cause” and “effect” hidden within malware execution flow. This study presents two novel contributions: the first empirical application of the S4 model for malware analysis, and a comprehensive comparison of its performance against other deep learning architectures, laying the stepping stone for future research in this new paradigm. 
\end{abstract}


\begin{figure*}[t]
    \centering
    \includegraphics[width=\textwidth]{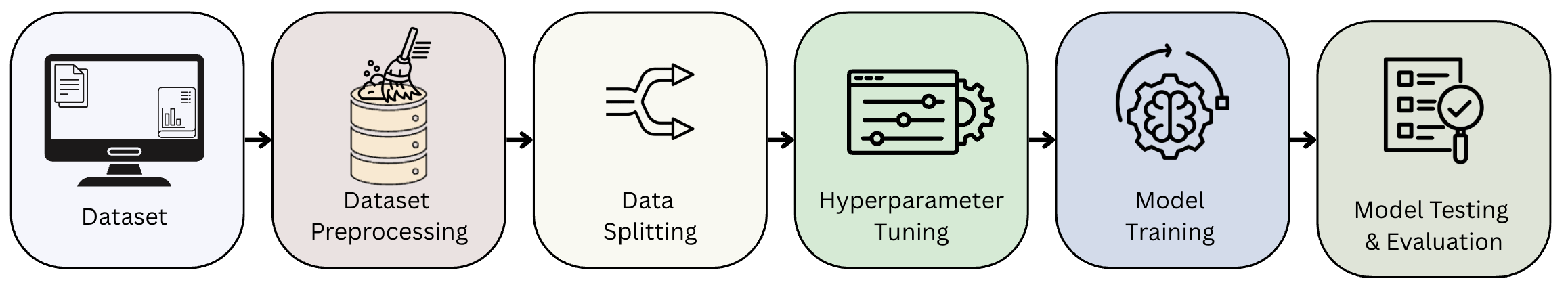}
\caption{Overall System Workflow}  
\vspace{-5mm}
 \label{fig:architecture}
\end{figure*}

\section{Introduction}
The gradual growth of AI, 5G, and edge computing is propelling the growth of the Internet of Everything (IoE), a network that connects everything and everyone. IoE is an evolution of IoT, and it faces the greatest threat to cybersecurity, malware. Malware does not target only professionals within the domain of cybersecurity, nor does it target a specific platform or operating system, such as embedded systems, smart cars, or mobile phones. Any device with Internet connectivity is exposed to different types of malware, such as viruses, worms, ransomware, trojan horses, and so on, each designed in a unique way to exploit vulnerabilities \cite{HistoryMalwareIBM2023}. With the projected increase in the number of devices on the Internet of 39 billion by 2030 \cite{noauthor_number_nodate}, this rise corresponds to a rapid increase in vulnerabilities. A growing challenge is imposed on current cybersecurity systems due to this constant increase in vulnerabilities, and, coincidentally, the complexity and scale with which malware variants are being created. 

Additionally, AV-Test roughly reports about 450,000 new malware variants per day [2], and according to a 2025 SonicWall report, malicious actors now utilize public exploit code within 48 hours of its release, leading to an increase in malware volume with a sharp rise in May 2024 at 92\% \cite{2025SonicWallCyber}. The report also highlights that IoT attacks increased by 124\% within that year \cite{2025SonicWallCyber}. In the United States, many critical sectors have been sitting ducks for ransomware throughout 2025 and 2026. The healthcare sector reported one of the biggest breaches, nearly affecting 193 million people, as well as disrupting services nationwide \cite{noauthor_change_nodate}, \cite{alder_largest_2026}. In 2025 alone, over 600 healthcare breaches were reported; likewise, attacks were launched on the education, finance, and government sectors, leading to downtime and data loss resulting in substantial operational and financial costs \cite{EscalatingRansomwareThreats}, \cite{ResourcesCISA}. These figures illustrate how costly malware attacks have become with this rapid growth rate. This complicates early detection and timely mitigation.

Traditionally, malware detection systems rely on detecting a unique digital footprint (malware signature) \cite{KeyMalwareDetection} or work by searching for commands not normally present in a benevolent application \cite{WhatHeuristicAnalysis}. Detection is performed by examining the signature or behavior of a program and comparing it to known threats. A software is flagged when it matches the signature of a known threat and may go unnoticed if it does not match a known threat. To address these limitations, anti-malware vendors increasingly adopted machine learning (ML) techniques to enhance detection and classification \cite{KasperskyRansomwareReport2025}, yet they remain limited in their ability to capture new malware variants at this rapid rate of growth. Recent research has shifted towards Deep Learning (DL) approaches to learn complex features from malware datasets for predictions, such as Convolutional and Recurrent Neural Networks (CNNs \& RNNs). The former is restricted in its ability to model long-term dependencies, and the latter, particularly Long Short-Term Memory (LSTM), becomes computationally expensive and suffers from gradient vanishing and exploding problems \cite{tuBridgingShortLongTerm2025}. While the introduction of Transformer models addresses this gap, they often do so at the cost of computation.

Malware classification research has taken several directions, with most work centered on feature extraction to improve multiclass performance. One line of research preserves raw binary structure during preprocessing, treating executables as one-dimensional byte signals. Wilkie et al. reported F1-scores of 0.503 and 0.507 for type and family classification on the large-scale MalNet dataset, indicating that low-level byte patterns retain enough information to support fine-grained family distinctions \cite{wilkieSignalBasedMalwareClassification2025}. Opcode-based approaches have progressed from n-gram models to architectures combining convolutional layers with bidirectional LSTM networks, reaching above 81\% accuracy across twenty families \cite{sainiOpCodeBasedMalwareClassification2025}, \cite{dangMalwareClassificationUsing2021}. Behavioral methods operating on API call sequences have shown that LSTM models consistently outperform nearest-neighbor, support vector machine, and decision tree baselines at the family level \cite{catakDeepLearningBased2020}. Graph-based models encoding directional API relationships have also demonstrated gains on datasets such as Cuckoo Sandbox, where structural dependencies provide useful signals for separating closely related families \cite{mGraphBasedMalware2025}. Portable executable metadata, packet payloads, and hybrid ensemble systems have produced strong results particularly when family count is small or class imbalance is addressed \cite{nguyenEfficientFrameworkMultiClass2025}, \cite{steinTransformerBasedFrameworkPayload2024}, \cite{cyptoIntelligentMalwareClassification2025}. Among these, random forests, gradient boosting, and SVMs applied directly to tabular PE features have shown competitive classification accuracy \cite{cyptoIntelligentMalwareClassification2025}, \cite{miraouiFrontiersBinaryMulticlass}, though shallow decision boundaries may limit performance as families grow structurally similar. Similarly, Andam et al. proposed a hybrid framework combining an Autoencoder Feature Extractor with a Model-Agnostic Meta-Learning classifier for few-shot malware detection, achieving up to 93.65\% accuracy in binary classification tasks \cite{11337394}. Overall, these studies show that multiclass malware classification improves when models can retain long-range structure and capture detailed dependencies, but many existing approaches remain limited by sequence length constraints or computational cost.

To address these pitfalls, this study examines the S4 model \cite{guMambaLinearTimeSequence2024}, \cite{guEfficientlyModelingLong2022} for binary malware detection and multiclass family classification. Unlike recurrent and attention-based architectures, S4 applies an explicit state space formulation that captures long-range dependencies at linear computational complexity. Malware samples are represented as sequential feature inputs and passed through stacked state space layers, enabling the model to recover global structural patterns that standard deep learning architectures tend to miss. Shared state representations learned through state space dynamics are used to jointly identify discriminative features across both tasks, with particular emphasis on family-level classification where subtle inter-variant differences make accurate discrimination more demanding. This serves as the basis for the proposed methodology in this study, and thus, the objectives of this study is summarized as follows:
\begin{enumerate}
\item We present the first empirical application of an S4 model to sequential malware analysis.
\item We design and implement an S4-based framework for binary detection and multiclass family classification.
\item We benchmark S4 against CNN, LSTM, GRU, and Transformer baselines, evaluating accuracy and computational efficiency.
\end{enumerate}
\begin{figure*}[t]
    \centering
    \includegraphics[width=\textwidth]{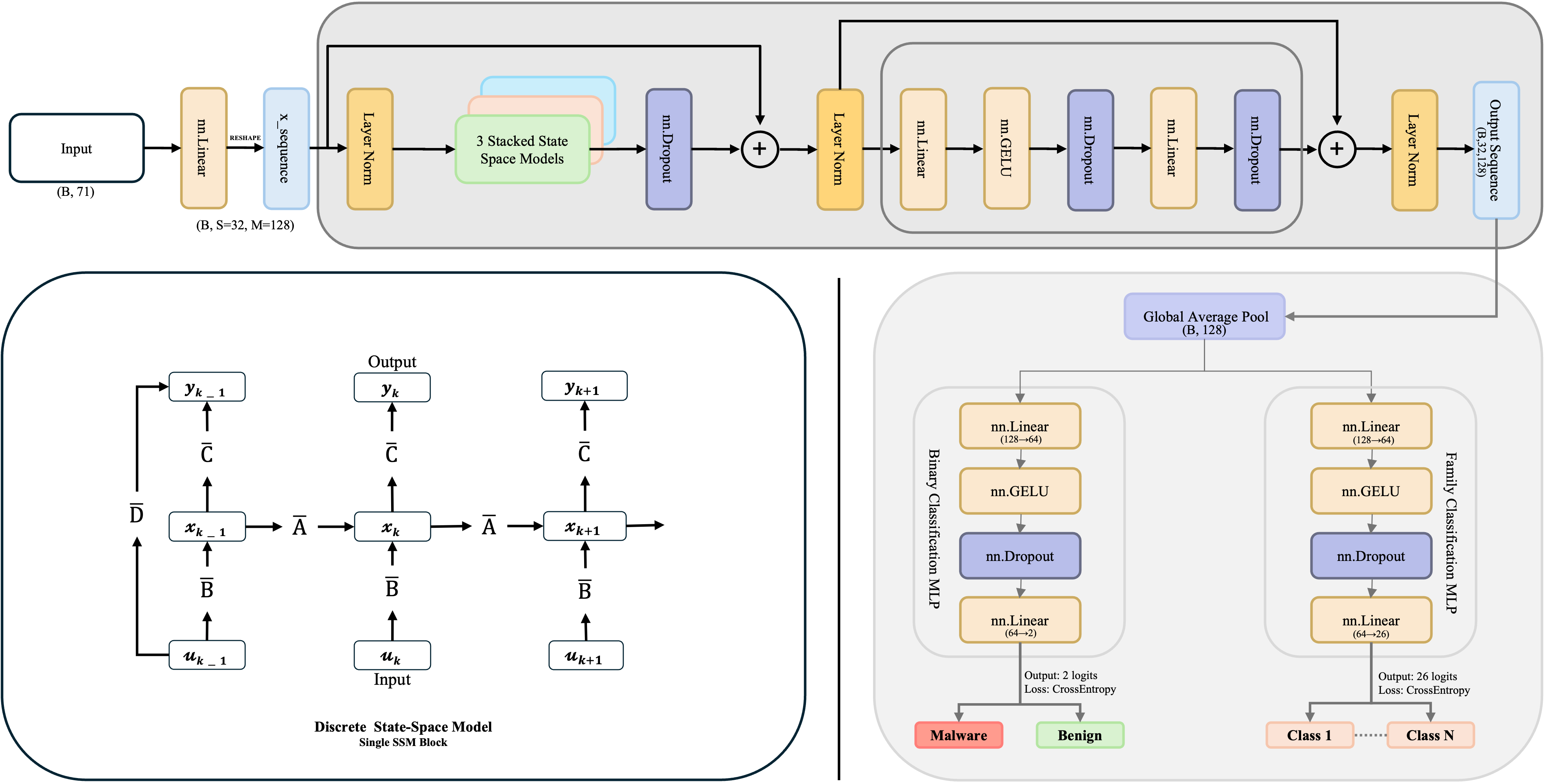}
\caption{Overview of the proposed state space based malware classification architecture, showing the sequence construction stage, stacked state space blocks with residual connections, and dual feed-forward classification networks. The discrete-time dynamics within a single state space block are shown in the lower-left panel.}  
\vspace{-5mm}
 \label{fig:ssm_architecture}
\end{figure*}

\section{Methodology}
This section entails the methodology used in this study, including the dataset and preprocessing. Following that, the proposed state-space-based (S4) model architecture is introduced, with a detailed description of the training steps, model evaluation, and hyperparameter tuning steps utilized.  The overall workflow of the proposed approach is illustrated in Fig. \ref{fig:architecture}.

\subsection{Dataset and Preprocessing}
This study uses the Ransomware Dataset 2024, publicly available on Zenodo ~\cite{amjadhussainRansomwareDataset20242024}. This dataset contains a total of 21,752 samples (10,876 malicious and 10,876 benign files), and this even distribution between malicious and benign samples minimizes bias for supervised learning. The dataset spans twenty-six malware families and provides static numerical features extracted from each file, describing structural properties of the executable. These features include metadata fields, section-related attributes, and characteristics that reflect the internal layout of the file. Non-numeric identifiers such as hash values and text-based labels are removed prior to preprocessing. All remaining features are converted to numeric form, and we performed feature normalization using Z-score, defined as

\begin{equation}
x_{\text{scaled}} = \frac{x - \mu}{\sigma},
\end{equation}
where $\mu$ and $\sigma$ represent the mean and standard deviation of each feature across the dataset. This normalization process prevents features with large magnitudes from dominating the learning process and improves optimization stability. Additionally, the malware family labels are encoded using integer label encoding to support multi-class classification. Since the state space model operates on sequential inputs, each static feature vector is projected into a learned sequence representation.  Specifically, the normalized feature vector $\mathbf{f} \in \mathbb{R}^{D}$, where $D=71$ denotes the number of retained features, is mapped to a sequence $\mathbf{U} \in \mathbb{R}^{L \times d_{\text{model}}}$ using a linear projection, where $L = 32$ and $d_{\text{model}} = 128$. Although the original data is not inherently sequential, this conversion is necessary because the S4 architecture is defined over sequences and applies its recurrent state updates along the sequence dimension. By projecting the feature vector into L positions, the model distributes its computation across multiple steps, enabling the state space layers to capture interactions among feature subgroups rather than treating the input as a single flat vector.

\subsection{Model Architecture}
The proposed S4 model is built around a discrete state space formula designed to capture global structure across static malware features. Instead of relying on local convolutional filters or attention-based pairwise interactions, the architecture processes each sample through recurrent state updates guided by learnable system dynamics. This allows information from all input attributes to be integrated over multiple computational steps. Fig. ~\ref{fig:ssm_architecture} presents an overview of the S4 model framework. Each input sample is first represented as a fixed-length feature vector and then transformed into a latent sequence that can be processed by stacked state space blocks. The final representation is pooled and passed to task-specific classifiers for malware detection and family classification.

At the core of each block is a linear time-invariant state space system. In continuous time, the system is defined as
\begin{equation}
\dot{\mathbf{x}}(t) = \mathbf{A}\mathbf{x}(t) + \mathbf{B}\mathbf{u}(t),
\end{equation}
\begin{equation}
\mathbf{y}(t) = \mathbf{C}\mathbf{x}(t) + \mathbf{D}\mathbf{u}(t),
\end{equation}
where $\mathbf{x}(t) \in \mathbb{R}^{d_{\text{state}}}$ denotes the internal state, $\mathbf{u}(t) \in \mathbb{R}^{d_{\text{model}}}$ is the input, and $\mathbf{y}(t) \in \mathbb{R}^{d_{\text{model}}}$ is the output. 
The matrices $\mathbf{A}$, $\mathbf{B}$, $\mathbf{C}$, and $\mathbf{D}$ are learned during training and control how information flows through the system.

For use in deep learning, the continuous formulation is discretized and applied over indexed sequence positions. 
The resulting discrete-time system is written as
\begin{equation}
\mathbf{x}_{k+1} = \mathbf{A}_{d}\mathbf{x}_{k} + \mathbf{B}_{d}\mathbf{u}_{k},
\end{equation}
\begin{equation}
\mathbf{y}_{k} = \mathbf{C}\mathbf{x}_{k} + \mathbf{D}\mathbf{u}_{k},
\end{equation}
The model updates an internal state recurrently as it processes the input sequence.
To handle the continuous-time formulation, the dynamics are discretized using a first-order forward Euler scheme, with 
$\mathbf{A}_{d} = \mathbf{I} + \Delta t\,\mathbf{A}$ and
$\mathbf{B}_{d} = \Delta t\,\mathbf{B}$,
The timesetp $\Delta t$ t is learned jointly with the other model parameters, which allows the scale of the state updates to adjust during training.
For implementation, each sample is represented using a fixed set of static features extracted from the dataset. 

After preprocessing, each sample is represented as a normalized feature vector with 71 elements. 
Algorithm~\ref{alg:ssm_forward} illustrates the complete forward computation of the proposed model and its implementation.
\begin{algorithm}[t]
\caption{Forward Pass of the State Space Malware Classification Model}
\label{alg:ssm_forward}
\begin{algorithmic}[1]
\REQUIRE Normalized feature matrix $\mathbf{X} \in \mathbb{R}^{B \times D}$
\ENSURE Detection logits $\boldsymbol{\ell}_{\text{det}}$, family logits $\boldsymbol{\ell}_{\text{fam}}$

\STATE $\mathbf{Z} \leftarrow \text{Linear}(\mathbf{X})$
\STATE $\mathbf{U} \leftarrow \text{reshape}(\mathbf{Z}) \in \mathbb{R}^{B \times L \times d_{\text{model}}}$
\STATE $\mathbf{U} \leftarrow \text{LayerNorm}(\mathbf{U})$

\FOR{$i = 1$ to $N_{\text{blocks}}$}
    \STATE $\mathbf{U}_{\text{res}} \leftarrow \mathbf{U}$
    \STATE $(\mathbf{Y}, \mathbf{x}_{\text{final}}) \leftarrow \text{StateSpaceLayer}(\mathbf{U})$
    \STATE $\mathbf{U} \leftarrow \text{LayerNorm}(\mathbf{Y} + \mathbf{U}_{\text{res}})$
    \STATE $\mathbf{U}_{\text{res}} \leftarrow \mathbf{U}$
    \STATE $\mathbf{U} \leftarrow \text{FFN}(\mathbf{U})$
    \STATE $\mathbf{U} \leftarrow \text{LayerNorm}(\mathbf{U} + \mathbf{U}_{\text{res}})$
\ENDFOR

\STATE $\mathbf{h} \leftarrow \text{GlobalAvgPool}(\mathbf{U})$
\STATE $\boldsymbol{\ell}_{\text{det}} \leftarrow \text{MLP}_{\text{det}}(\mathbf{h})$
\STATE $\boldsymbol{\ell}_{\text{fam}} \leftarrow \text{MLP}_{\text{fam}}(\mathbf{h})$

\RETURN $\boldsymbol{\ell}_{\text{det}}, \boldsymbol{\ell}_{\text{fam}}$
\end{algorithmic}
\end{algorithm}

The feature vector is first passed through a linear projection and reshaped into a latent sequence of length $L = 32$, where each element has dimension $d_{\text{model}} = 128$. 
This step does not impose any temporal meaning on the features; instead, it provides a structured representation that allows the state space layers to apply repeated state updates over several computational steps.

The resulting sequence is processed by a stack of three identical state space blocks. 
Each block contains a single discrete-time state space layer, followed by residual connections, layer normalization, and a position-wise feed-forward network. 
Within each block, the state space layer iterates over the sequence dimension and produces an output sequence of the same shape as the input. 
Residual connections are applied around both the state space layer and the feed-forward network to maintain stable signal propagation across blocks.

The feed-forward network consists of two fully connected layers with a GELU activation function. 
The first layer expands the latent dimension to $2d_{\text{model}}$, and the second projects it back to $d_{\text{model}}$. 
Dropout is applied after the state space layer and after the feed-forward network for regularization.

After the final state space block, the sequence output is aggregated using global average pooling along the sequence dimension. 
This produces a fixed-length representation of dimension $d_{\text{model}}$, which is shared by two task-specific classifiers. 
One classifier performs binary malware detection, while the second predicts the malware family. 
Both classifiers are implemented as small feed-forward networks and output unnormalized logits, which are trained using cross-entropy loss.

The architectural configuration and dimensional settings used in all experiments are summarized in Table~\ref{tab:architecture}.
\begin{table}[t]
\centering
\caption{Configuration of the proposed state space based model}
\label{tab:architecture}
\begin{tabular}{l c}
\hline
Component & Setting \\
\hline
Number of input features ($D$) & 71 \\
Sequence length ($L$) & 32 \\
Model dimension ($d_{\text{model}}$) & 128 \\
State dimension ($d_{\text{state}}$) & 64 \\
Number of state space blocks & 3 \\
State space layers per block & 1 \\
Feed-forward network depth & 2 layers \\
Feed-forward expansion ratio & $2 \times d_{\text{model}}$ \\
Activation function & GELU \\
Normalization & LayerNorm \\
Residual connections & Around SSM and FFN \\
Dropout rate & 0.1 \\
Pooling method & Global average pooling \\
\hline
\end{tabular}
\end{table}
\begin{figure*}[b]
    \centering
    \includegraphics[width=\textwidth]{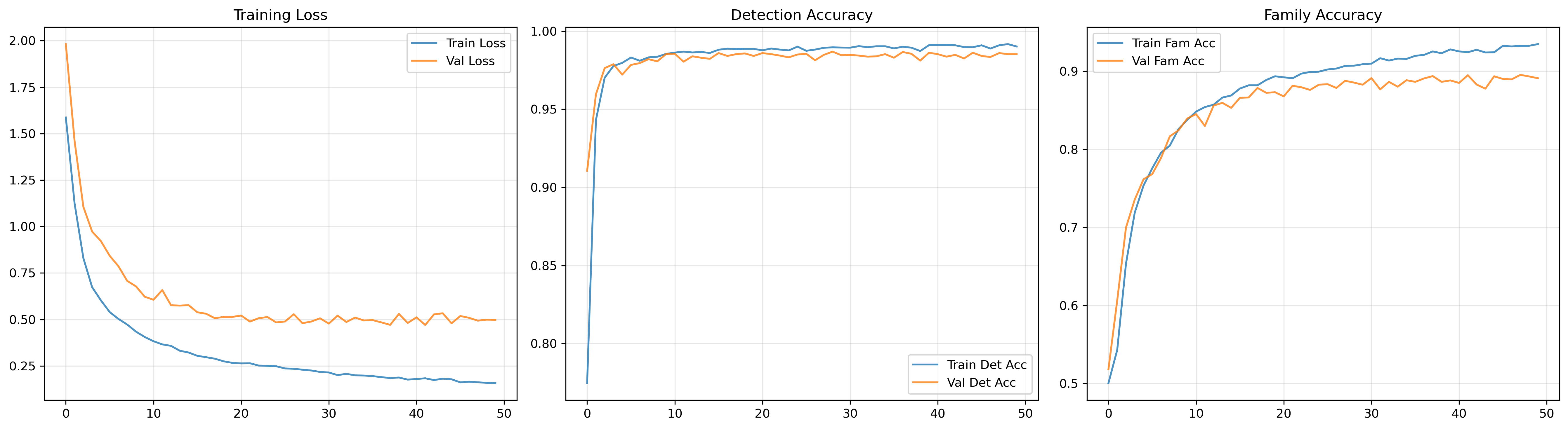}
    \caption{Training and validation loss, binary detection accuracy, and malware family classification accuracy over 50 training epochs.}
    \vspace{-5mm}
    \label{fig:training_dynamics}
\end{figure*}
\subsection{Training and Hyperparameter Tuning}
In this study all models were trained and evaluated using a fixed train–validation–test split derived from the full dataset. The dataset is partitioned using stratified sampling based on the binary detection labels to preserve the malicious–benign class balance. Specifically, 60\% of the samples were used for training, while the remaining 40\% were split evenly into validation (20\%) and test (20\%) sets. Both splits were performed using a fixed random state of 42 to ensure reproducibility of the data partitioning.
Model training was performed using the AdamW optimizer with a learning rate of 0.001 and a weight decay of 0.01. Training was conducted for a fixed number of 50 epochs with a batch size of 256. Gradient norms were clipped to a maximum value of 1.0 to stabilize optimization. Dropout with a rate of 0.1 was applied throughout the model, including within the state space blocks and the classification heads. A cosine annealing learning rate scheduler was applied during training to gradually allow the learning rate to decay and stabilize the optimization process as the model converged. The model is designed for two objectives: binary malware detection and malware family classification. As such, both objectives use the cross-entropy loss. During training, the total loss is computed as a weighted sum using:

\begin{equation}
\mathcal{L}_{\text{train}} = 0.3\,\mathcal{L}_{\text{det}} + 0.7\,\mathcal{L}_{\text{fam}},
\end{equation}
where $\mathcal{L}_{\text{det}}$ and $\mathcal{L}_{\text{fam}}$ denote the detection and family classification losses, respectively. This prioritizes the family classification, which tells us the kind of malware the sample is. 
During validation and testing, losses are reported as an unweighted sum of the two components. We used validation family classification accuracy as the sole metric for model checkpointing. During training, we only saved the model version that reached the highest accuracy on this task to use for the final evaluation. All experiments were conducted on a single device to ensure a consistent environment to compare the S4 model against the deep learning baselines. 

\begin{table}[t]
\centering
\caption{Hyperparameter search space used for grid search tuning}
\label{tab:grid}
\begin{tabular}{l c}
\hline
Parameter & Values \\
\hline
Model dimension ($d_{\text{model}}$) & \{128, 256\} \\
State dimension ($d_{\text{state}}$) & \{32, 64\} \\
Number of state space blocks & \{2, 3\} \\
Learning rate & \{1e{-4}, 5e{-4}\} \\
Family loss weight & \{0.5, 0.7\} \\
Dropout rate & \{0.1, 0.2\} \\
Batch size & \{128, 256\} \\
\hline
\end{tabular}
\end{table}
\paragraph{Hyperparameter Tuning}
Hyperparameter tuning was conducted using an explicit grid search strategy for all models. For S4, the search space covered architectural, optimization, and regularization parameters, as summarized in Table~\ref{tab:grid}. For the baseline models, the search was restricted to learning rate, batch size, and dropout rate across the same candidate values, with architecture-specific parameters such as hidden size and number of layers varied over two to three values. In all cases, 32 trials of 10-epoch runs were evaluated and the best configuration per model was selected by validation family accuracy before retraining for 50 epochs. The detection loss weight was fixed at 0.3, while the family classification loss weight was varied as part of the grid.

Model selection during tuning was performed using validation family classification accuracy. After identifying the best-performing configuration, the model was retrained from scratch using the full training set for 50 epochs. The final model checkpoint was again selected using the highest validation family accuracy, and the final evaluation was performed on the held-out (unseen) test set.

\subsection{Evaluation Metrics}
\label{subsec:metrics}
To evaluate the performance of the proposed S4 model, we used several metrics to evaluate the malware classification framework and capture its performance, firstly, on detecting malware or benign. Second, we look at per-class performance to see if the model is biased and identifies some families better than others. Finally, we use macro and weighted averages to summarize the results across the entire dataset. The difference between these two averages shows whether the model is biased. The macro average treats every malware class as equal, particularly focusing on the minority class. On the other hand, the weighted score scales the results based on the number of samples or true instances, and this in turn, favors the majority classes. Together, these metrics provide a comprehensive and balanced evaluation of the proposed model.
A summary of the evaluation metrics used in this study is provided in 
Table~\ref{tab:metrics}.

\begin{table}[t]
\centering
\caption{Evaluation metrics used for malware classification}
\label{tab:metrics}
\small
\setlength{\tabcolsep}{4pt}
\begin{tabular}{l c}
\hline
Metric & Formula \\
\hline
Accuracy &
$\dfrac{TP + TN}{TP + TN + FP + FN}$ \\

Precision &
$\dfrac{TP}{TP + FP}$ \\

Recall &
$\dfrac{TP}{TP + FN}$ \\

F1-score &
$\dfrac{2 \cdot \text{Precision} \cdot \text{Recall}}{\text{Precision} + \text{Recall}}$ \\

Macro-average &
$\dfrac{1}{C}\sum_{i=1}^{C} m_i$ \\

Weighted-average &
$\sum_{i=1}^{C} w_i m_i$ \\
\hline
\end{tabular}
\normalsize
\end{table}

\section{Results and Discussion}
In this section, we present the experimental results obtained and discuss the S4 model’s performance to CNN, LSTM, GRU, and Transformer using the evaluation metrics defined in Section~\ref{subsec:metrics}.The experiments were conducted on a Dell Precision 3260 workstation equipped with a 12th Gen Intel(R) Core(TM) i7-12700 processor and 32 GB of RAM, with GPU acceleration utilized when available, and on an independent test set out during the data preprocessing steps for evaluation (unseen data). We begin by assessing the training behavior, then move on to a detailed evaluation of the models’ performances. 

\subsection{Training}
In Fig. ~\ref{fig:training_dynamics}, we highlight the training loss, binary detection accuracy, and per-class classification of the proposed model over 50 epochs. The training and validation loss decrease sharply within the first few epochs and gradually stabilize as training progresses. The detection accuracy, on the other hand, rises sharply within a few epochs and remains fairly consistent thereafter. Similarly, the per-family accuracy increases; however, it’s a steady increase over the training epochs and stabilizes towards the later phase. Overall, from the results, we find that the model convergence is stable over the course of the training.  

\subsection{S4 Model Performance}
\begin{figure}[t]
    \centering
    \includegraphics[width=\columnwidth]{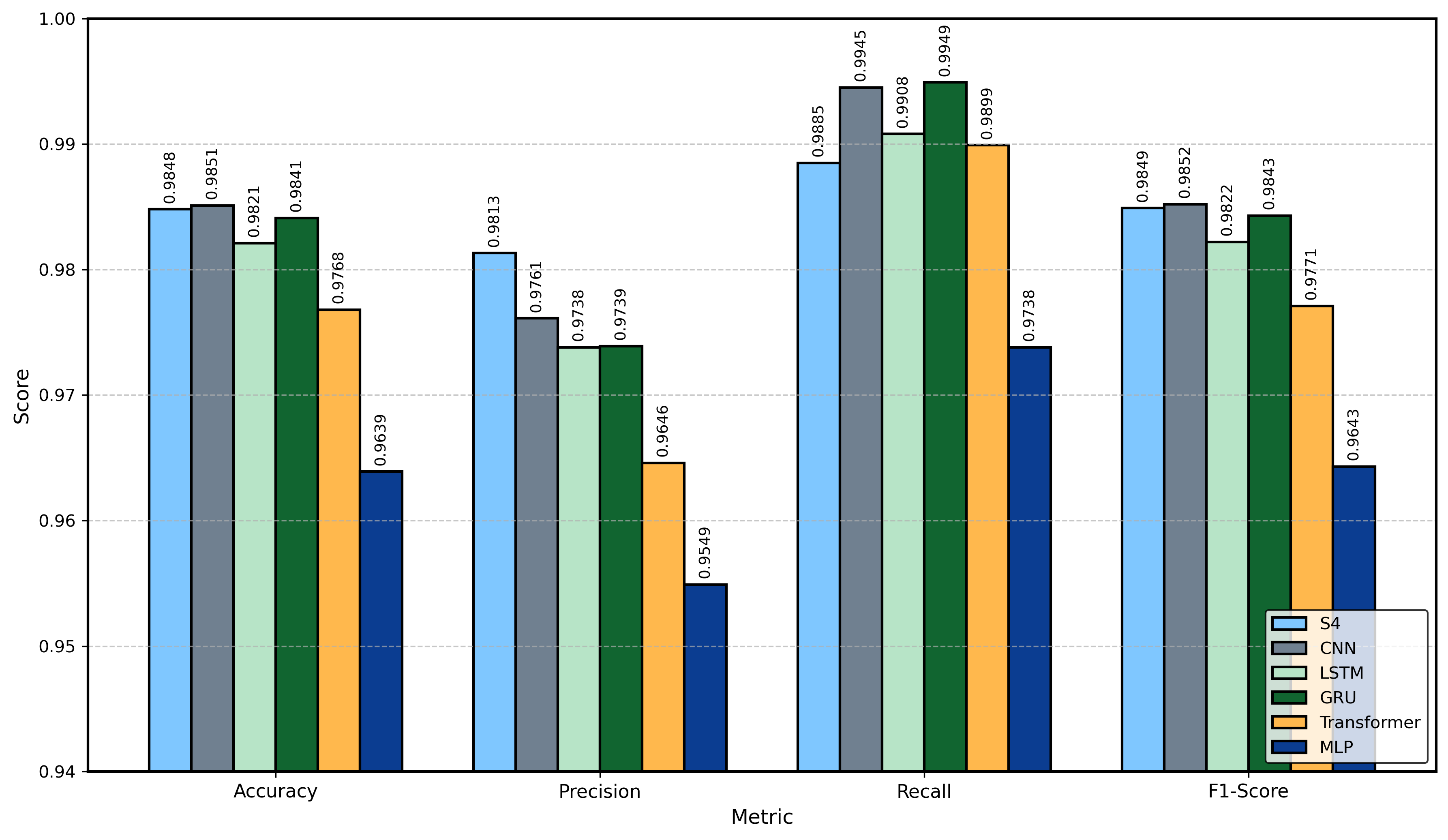}
    \caption{Binary malware detection performance across evaluated models in terms of accuracy, precision, recall, and F1-score.}
    \vspace{-5mm}
    \label{fig:binary_detection}
\end{figure}
\begin{figure}[t]
    \centering
    \includegraphics[width=\columnwidth]{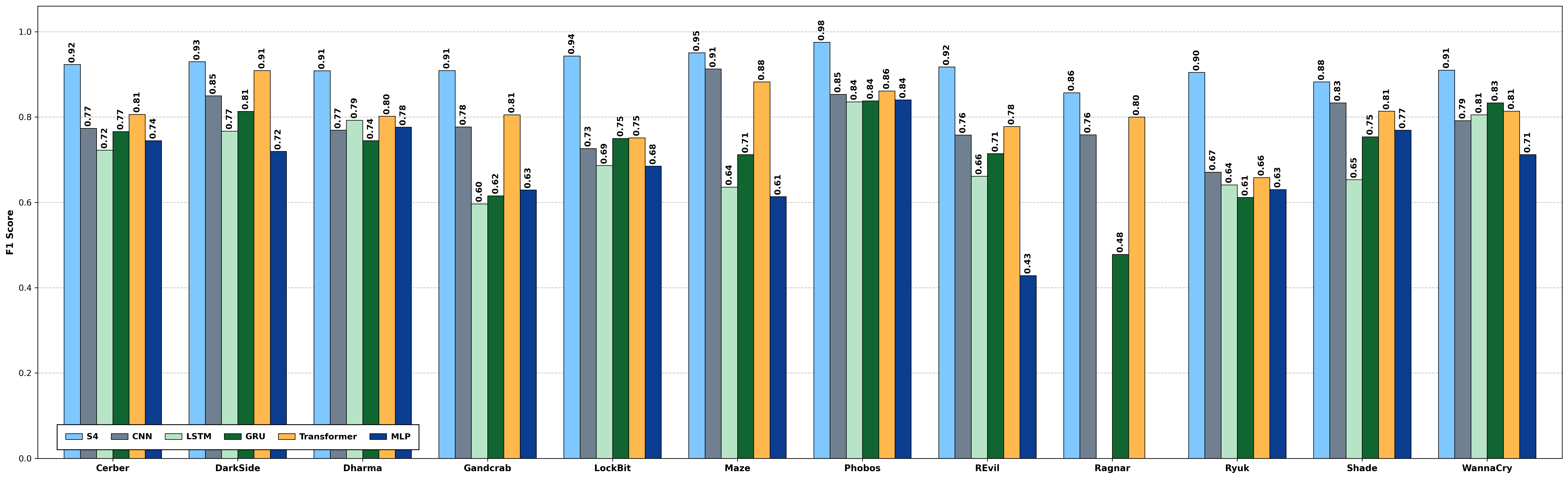}
    \caption{Per-class F1-scores for the selected ransomware families across different model architectures.}
    \vspace{-5mm}
    \label{fig:family_f1}
\end{figure}
\begin{figure}[b]
    \centering
    \includegraphics[width=\columnwidth]{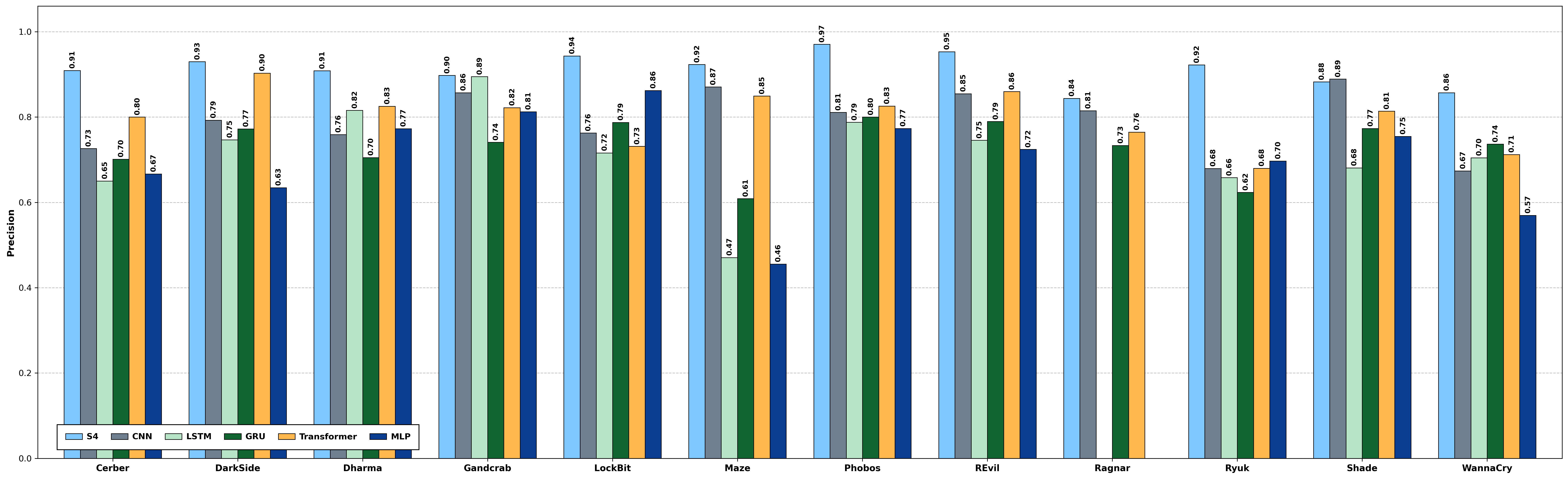}
    \caption{Per-class precision for the selected ransomware families across different model architectures.}
    \vspace{-5mm}
    \label{fig:family_precision}
\end{figure}
\begin{figure}[b]
    \centering
    \includegraphics[width=\columnwidth]{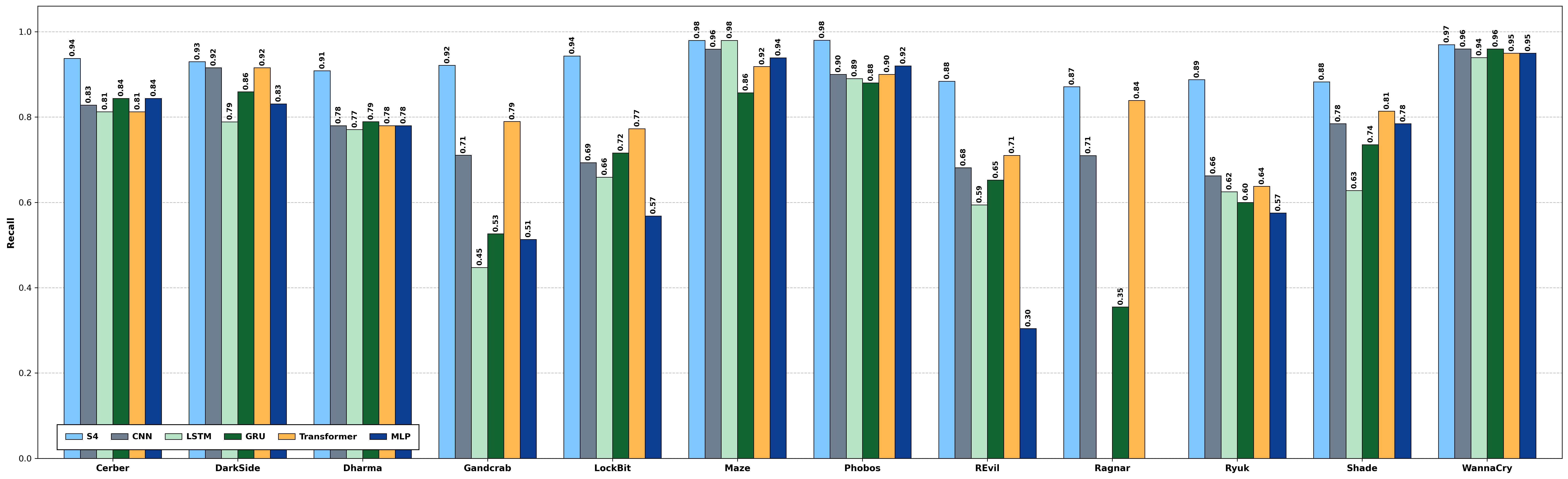}
    \caption{Per-class recall for the selected ransomware families across different model architectures.}
    \vspace{-5mm}
    \label{fig:family_recall}
\end{figure}
\begin{figure*}[t]
    \centering

    \begin{subfigure}{0.32\textwidth}
        \centering
        \includegraphics[width=\linewidth]{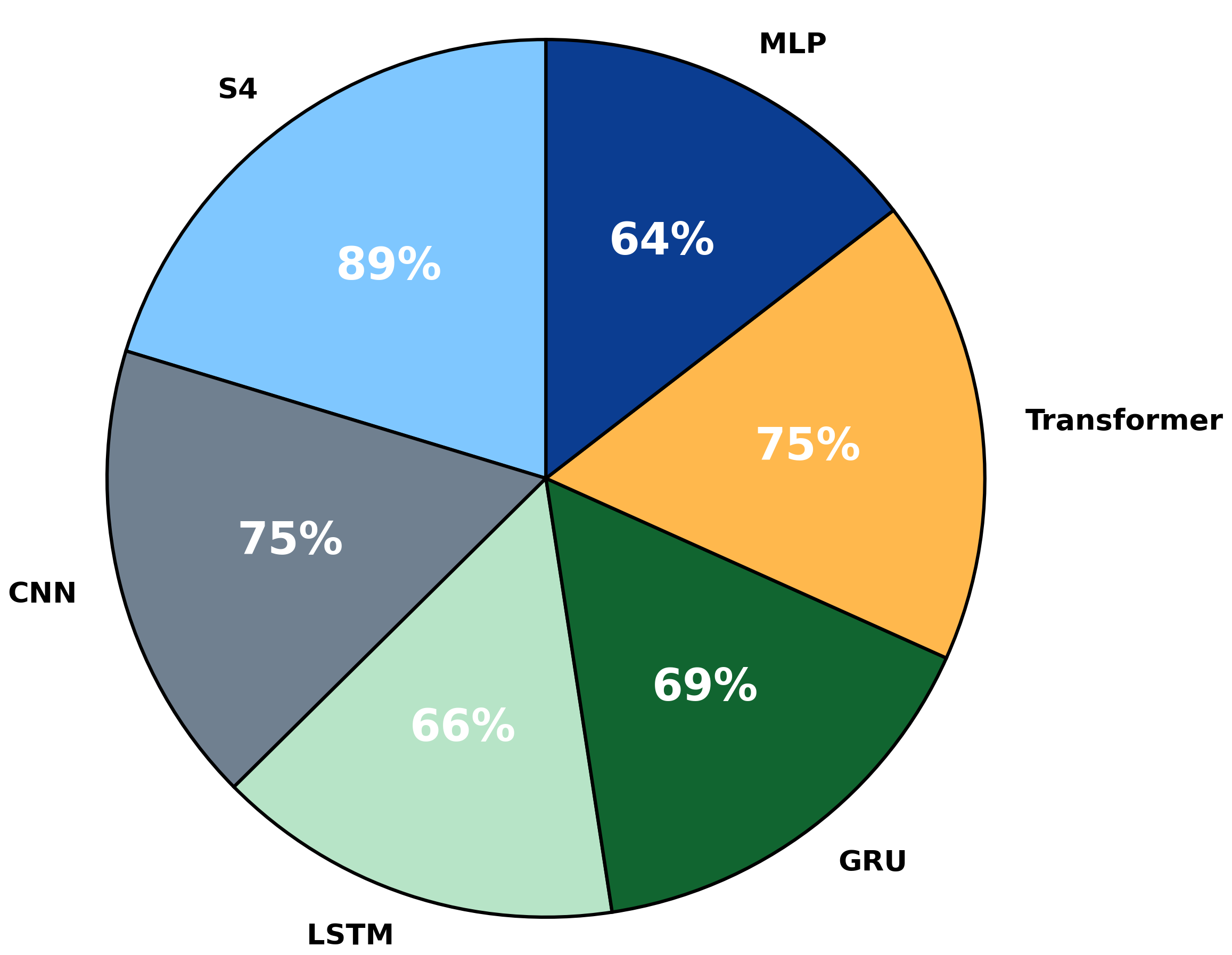}
        \caption{F1-score (macro)}
    \end{subfigure}
    \hfill
    \hfill
    \begin{subfigure}{0.32\textwidth}
        \centering
        \includegraphics[width=\linewidth]{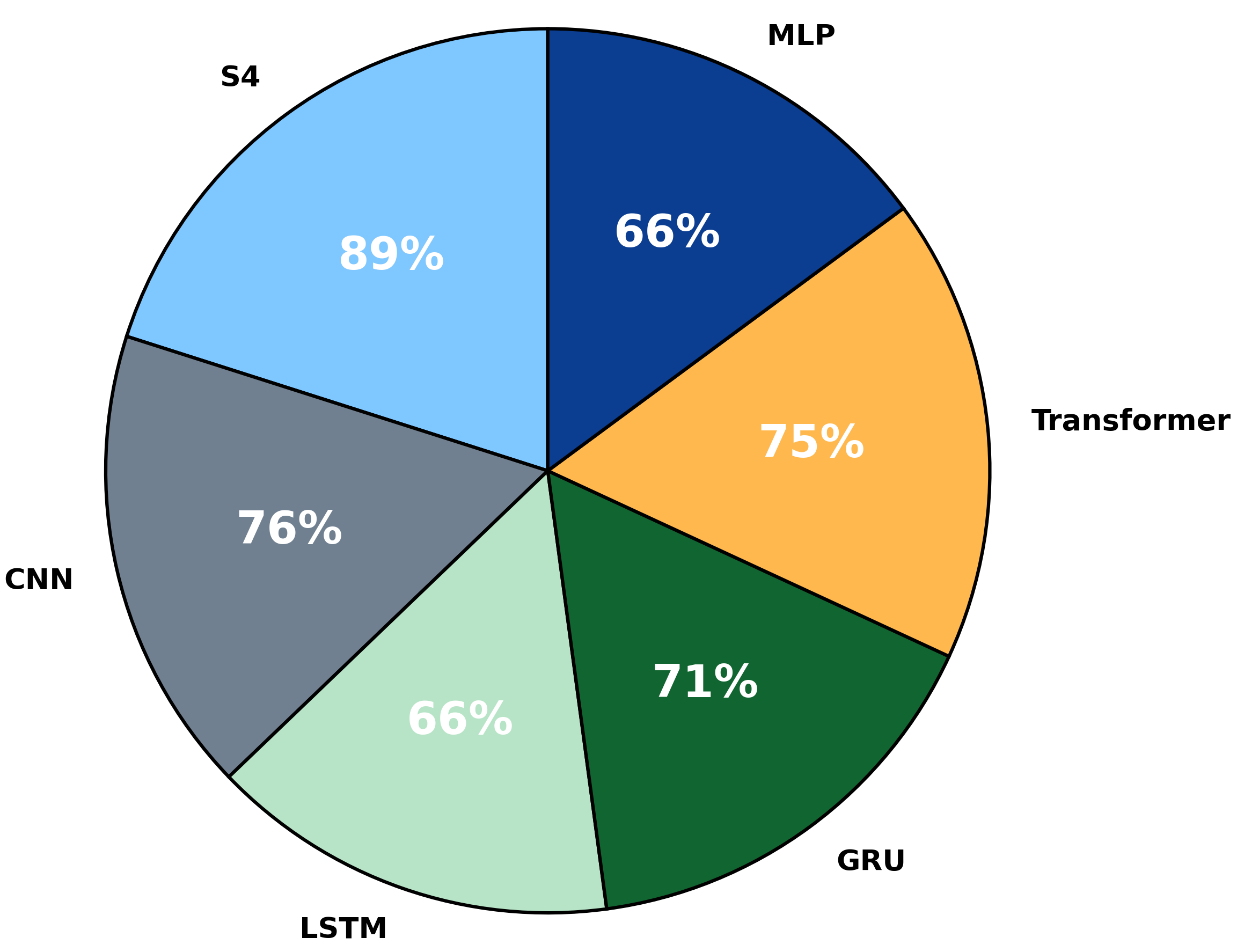}
        \caption{Precision (macro)}
    \end{subfigure}
    \hfill
    \begin{subfigure}{0.32\textwidth}
        \centering
        \includegraphics[width=\linewidth]{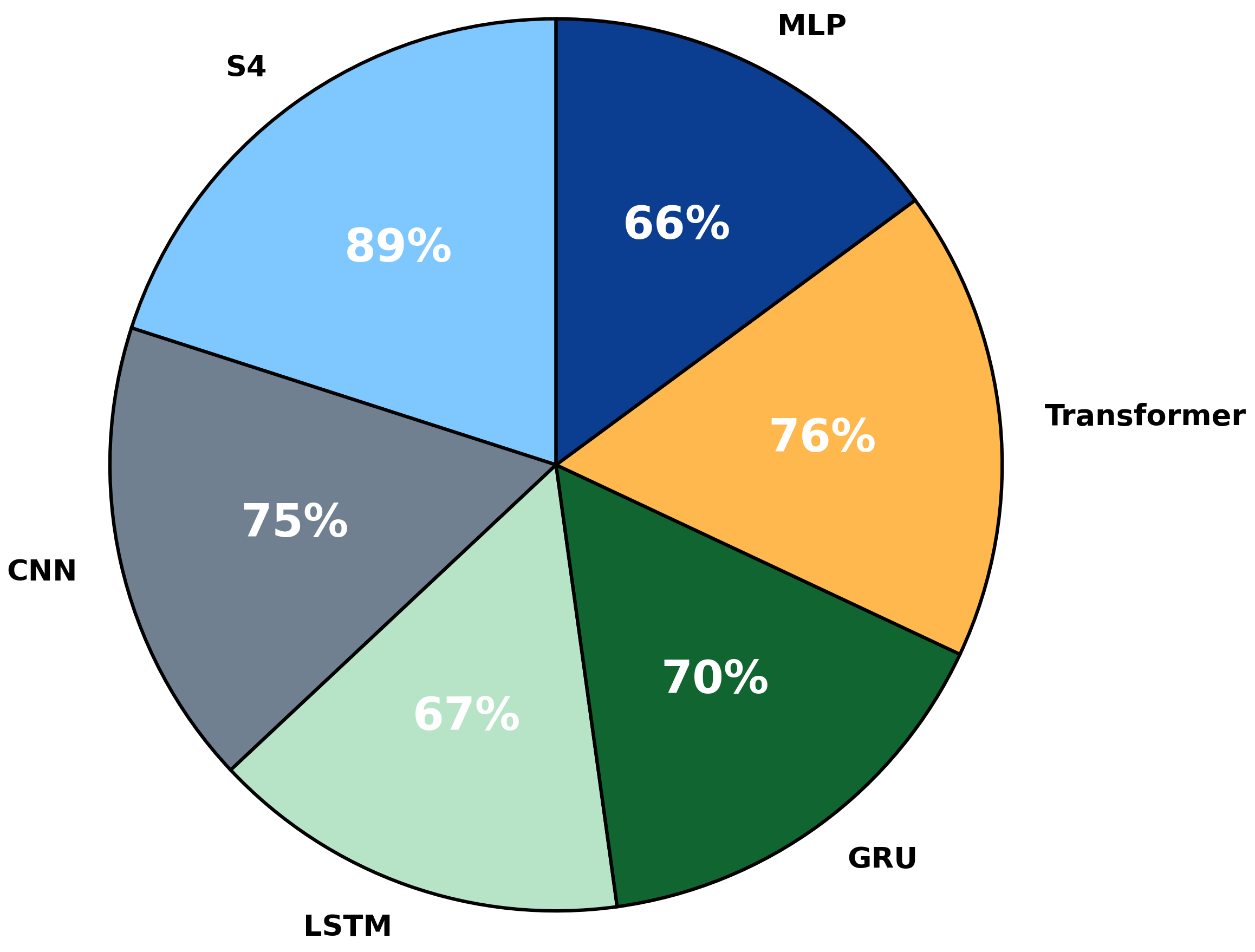}
        \caption{Recall (macro)}
    \end{subfigure}

    \vspace{0.6em}

    \begin{subfigure}{0.32\textwidth}
        \centering
        \includegraphics[width=\linewidth]{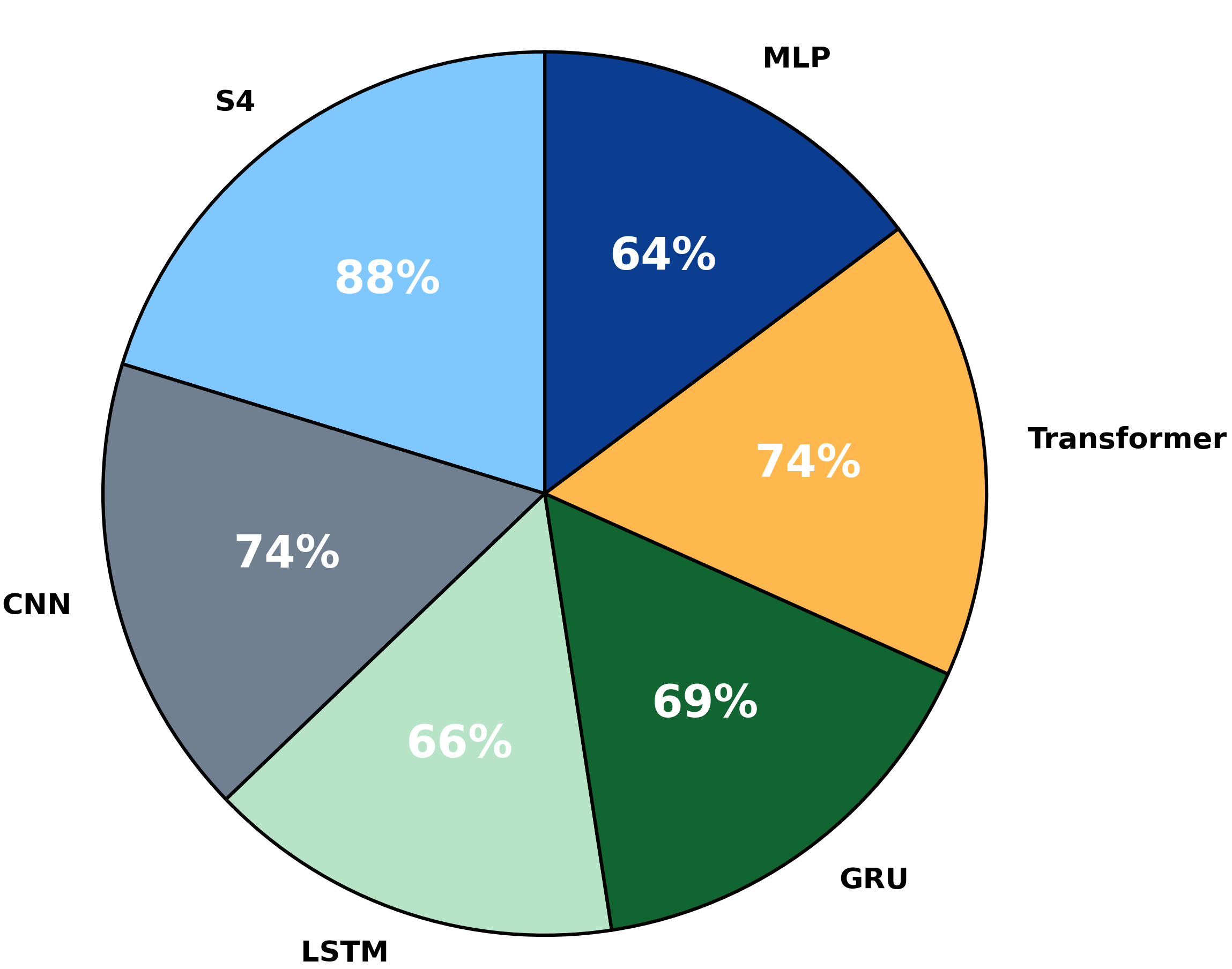}
        \caption{F1-score (weighted)}
    \end{subfigure}
    \begin{subfigure}{0.32\textwidth}
        \centering
        \includegraphics[width=\linewidth]{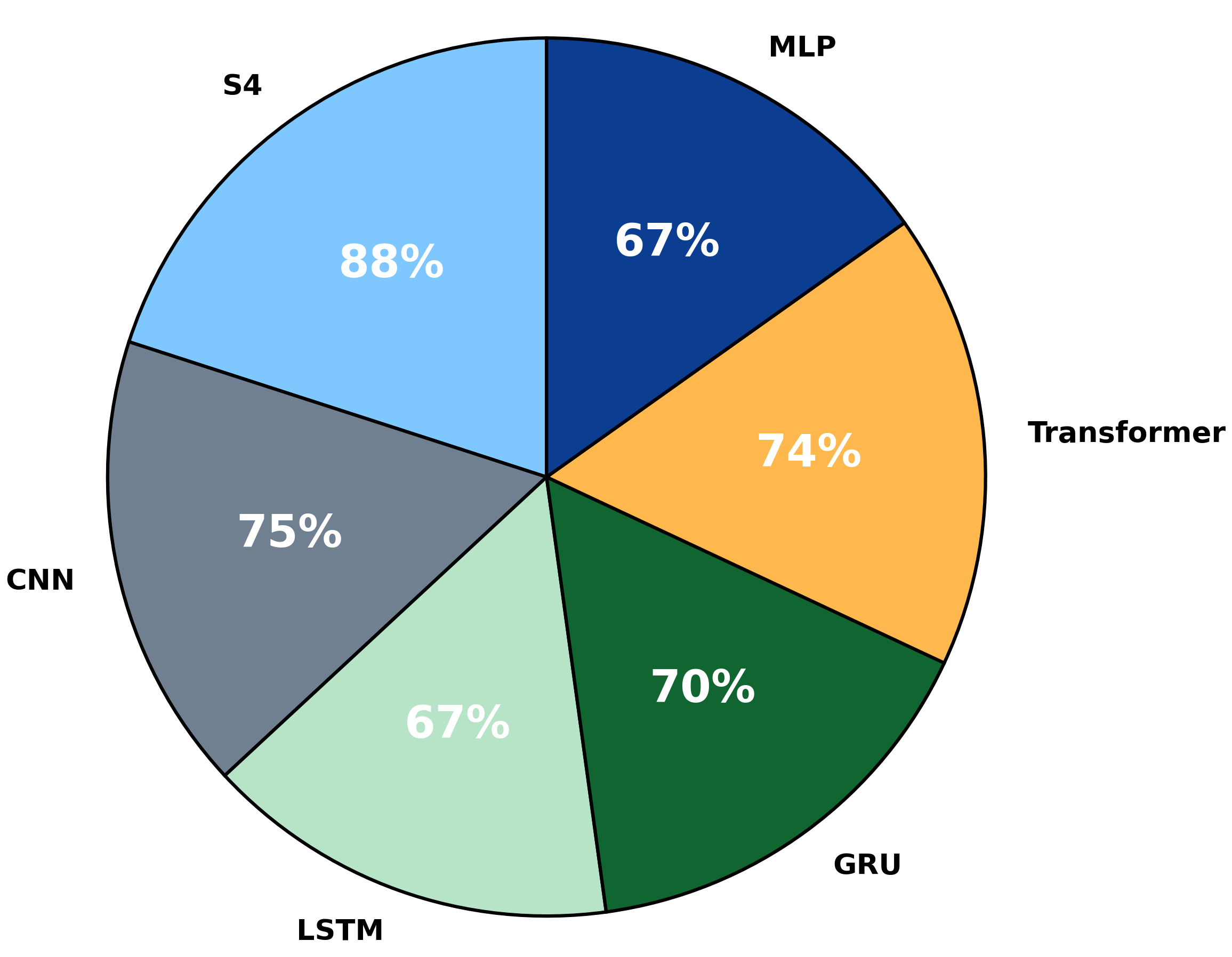}
        \caption{Precision (weighted)}
    \end{subfigure}
    \hfill
    \begin{subfigure}{0.32\textwidth}
        \centering
        \includegraphics[width=\linewidth]{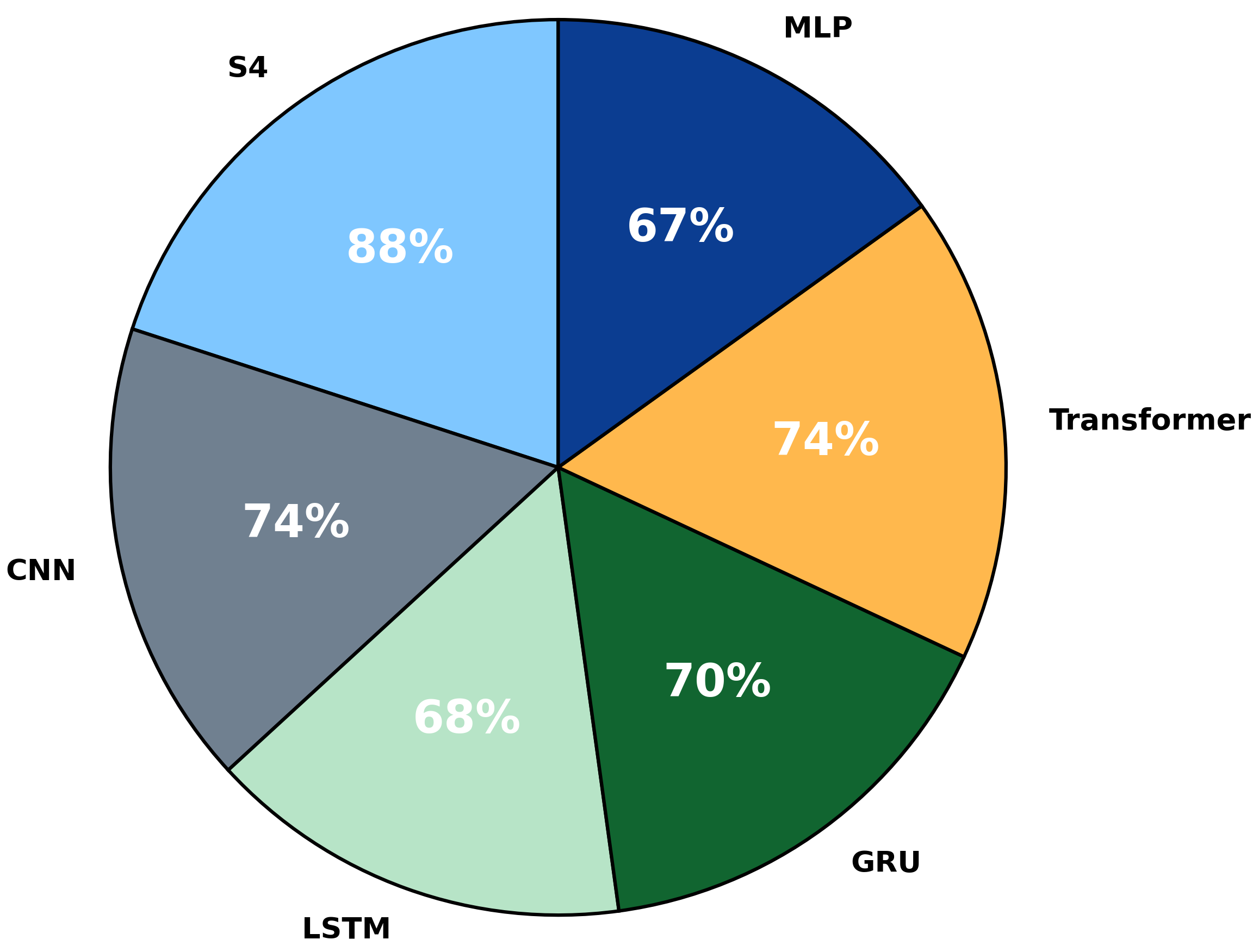}
        \caption{Recall (weighted)}
    \end{subfigure}

    \caption{Summary of macro-averaged and weighted performance metrics for malware family classification across evaluated models.}
    \vspace{-5mm}
    \label{fig:summary_metrics}
\end{figure*}
To evaluate the performance of the proposed S4 model, we first examine how it compares with other models on detecting malware, that is, whether the model can tell apart malware from benign. These results are illustrated in Fig. ~\ref{fig:binary_detection} and across all models, detection performance is consistently high with accuracy scores above 97\% and few discrepancies between the individual models. However, it is worth noting that the S4 model’s accuracy of 98.5\%, with precision, recall, and F1-score all exceeding 98\%, which is comparable to the other models considered, and strongly suggests that indeed the S4 model, after modeling the malware samples as long sequences, captures the necessary features to accurately detect malware. The remaining question, therefore, is not whether a sample can be detected as malicious, but how well each model can distinguish between different malware families. For this reason, we proceed to examine each model’s performance when classifying the detected malware.

Following the binary detection results, we evaluate how well each model distinguishes between different malware families.
The dataset contains samples from 26 malware families; however, to allow for a focused and interpretable analysis, we restrict attention to 12 ransomware families that are among the most prevalent and impactful in real-world attacks.
These include Cerber, DarkSide, Dharma, GandCrab, LockBit, Maze, Phobos, REvil, Ragnar Locker, Ryuk, Shade, and WannaCry. Fig.~\ref{fig:family_f1} presents the per-family F1-scores across all evaluated models. The S4 model achieves an F1-score of 0.90 or above in 10 of the 12 families, with peak scores recorded for Phobos (0.98), Maze (0.95), and LockBit (0.94), and no substantial degradation observed across the remaining families.

The baseline architectures exhibit considerably greater variance across families. The Transformer and CNN models remain competitive on families such as DarkSide and Maze but show marked deterioration on others. The MLP produces an F1-score of 0.43 on REvil and fails to generate a score for Ragnar, a result shared by several other baselines, indicating limited capacity to capture the discriminative patterns of these variants. LSTM and GRU show a comparable decline on Ryuk and Shade, where scores fall within the 0.60 to 0.75 range.
In these more challenging cases, S4 outperforms the baselines by margins of 15 to 20 percentage points, supporting the conclusion that the model's ability to capture long-range dependencies and subtle structural differences accounts for its advantage over conventional architectures.
Fig.~\ref{fig:family_precision} and Fig.~\ref{fig:family_recall} present per-family precision and recall scores. S4 maintains a consistent balance between the two metrics across families, which accounts for its strong F1-scores throughout. On Phobos, GandCrab, and LockBit, both precision and recall remain high, reflecting reliable identification with few false positives or missed detections.
Several baseline models show pronounced asymmetry between the two metrics. On REvil, Ragnar, and Shade, recall drops sharply even where precision holds relatively steady, suggesting these architectures are more prone to missing samples than to misclassifying them. The CNN on REvil is a clear example, recording 0.85 precision alongside 0.68 recall. The MLP shows the most severe version of this pattern, with simultaneous collapse in both metrics across multiple families, particularly REvil and Maze.
Fig.~\ref{fig:summary_metrics} summarizes aggregate performance using macro and weighted precision, recall, and F1-scores. Macro averages treat each family equally regardless of sample count, while weighted averages reflect class prevalence. S4 records a macro F1-score of 89\% and a weighted F1-score of 88\% across both aggregations, sitting substantially above the next closest models. The Transformer and CNN each achieve macro and weighted F1-scores of 75\% and 74\% respectively, while the MLP records the lowest scores overall, ranging from 64\% to 67\% across all aggregate metrics.
These results indicate that the performance advantage of S4 is not driven by a small number of well-represented families but holds across the full range of ransomware classes in the dataset. All reported metrics reflect a single experimental run; future work will report mean and standard deviation across multiple runs to quantify variance.

\section{Conclusion}
This paper evaluated a Structured State-Space Sequence (S4) model for binary malware detection and multi-class ransomware family classification using static features, and compared it with CNN, LSTM, GRU, Transformer, and MLP baselines. All models performed reliably on binary detection, recording accuracies above 97\%, with S4 reaching 98.5\%. At this task, the separation between benign and malicious samples was consistent across all architectures.
Notable differences became apparent at the family level. S4 achieved a macro F1-score of 89\% and a weighted F1-score of 88\%, while the CNN, LSTM, GRU, and Transformer baselines ranged between 75\% and 76\%, and the MLP did not reach 67\%. Precision and recall were also more evenly balanced within individual families; on Phobos, S4 recorded 0.97 precision and 0.98 recall, a consistency that the baseline models did not maintain across closely related variants.
Overall, the results indicate that structured state-space modeling is well suited for sequential malware analysis and improves family-level classification while maintaining strong detection performance.  This study is limited to a single dataset; future work will evaluate generalizability on additional benchmarks such as CIC-MalMem-2022 and BODMAS, which cover different feature modalities and temporal distribution shift. Additional directions include dynamic behavioral features such as API call traces, and robustness under concept drift as malware families evolve.

\bibliography{references}

\end{document}